\documentclass[12pt]{article} 
\usepackage{nips15submit_e,times}
\usepackage{hyperref}
\usepackage{url}
\usepackage{amsmath,amssymb}
\usepackage{enumerate}
\usepackage{graphicx}

\title{Multiversion Altruistic Locking}

\author{
Chinmay Chandak, Hrishikesh Vaidya, Sathya Peri\\
\texttt{\{cs13b1011,cs13b1035,sathya\_p\}@iith.ac.in} \\
Indian Institute of Technology, Hyderabad
}

\nipsfinalcopy 

\begin{document}

\maketitle

\vspace*{1cm}
\begin{abstract}
\emph{This paper builds on altruistic locking which is an extension of 2PL. It allows more relaxed rules as compared to 2PL. But altruistic locking too enforces some rules which disallow some valid schedules (present in VSR and CSR) to be passed by AL. This paper proposes a multiversion variant of AL which solves this problem. The report also discusses the relationship or comparison between different protocols such as MAL and MV2PL, MAL and AL, MAL and 2PL and so on. This paper also discusses the caveats involved in MAL and where it lies in the Venn diagram of multiversion serializable schedule protocols. Finally, the possible use of MAL in hybrid protocols and the parameters involved in making MAL successful are discussed.}
\end{abstract}
\vspace*{1cm}

\section{Motivation}  

\includegraphics[width=15cm,height=5cm]{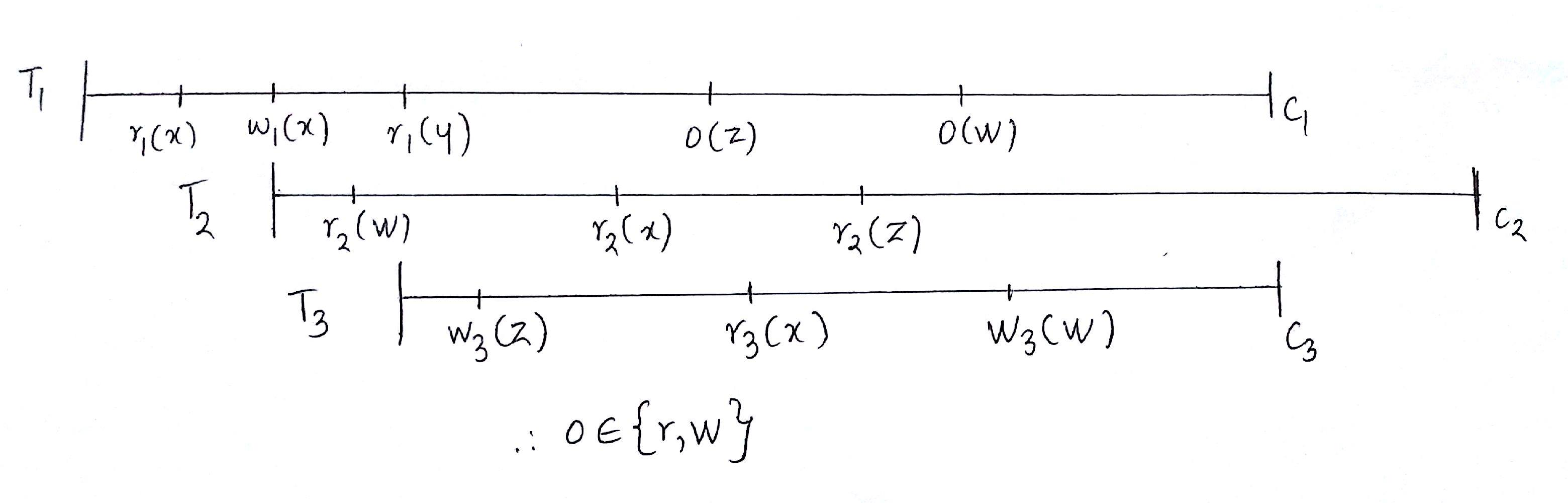}

Suppose $T_1$ is a long transaction with data operations on various variables while $T_2$ and $T_3$ are short transactions which just want to read the value of $x$. In 2PL, we saw that the \texttt{RO} or likewise transactions (here $T_2$ and $T_3$) suffer from time-lag until $T_1$ starts to unlock locks on $x$. AL resolved this problem to an extent in which once $T_1$ is done with $x$, it donates the lock to $T_2$ and $T_2$ reads the current version of x and similarly for $T_3$. By current, most recent committed version is implied (here $x_0$).
Now, if no further writes on $x$ take place, the write of $T_1$ on $x$ is useless since $T_2$ and $T_3$ read from $T_0$. If we know that $T_1$ will not abort, we can read from uncommitted versions as well i.e. $T_2$ and $T_3$ turn by turn can read either from $x_0$ or from $x_1$ if versions are assigned. Hence more usefulness in terms of garbage collection and donation of locks is seen with a multiversion variant.

\section{Multiversion Altruistic Locking}
\subsection{Definition}
The notion of a multiversion variant of altruistic locking can be seen from the motivation provided above. From now on, we'll abbreviate this protocol as MAL.\\
The key point in this protocol like AL would be donation of locks. Like AL, locks would be donated on variables but now since read operations have multiple choices of versions to read from, the field of conflicts (now multiversion) would be less and thus would allow more concurrency than AL; its single-version counterpart protocol.
\subsection{Rules}

The first three rules would be similar to AL of course.\\
\texttt{MAL1}: Items cannot be read or written by $t_i$ once it has donated them; that is, if $d_i(x)$ and $o_i(x)$ occur in a schedule $s,o \in {r, w}$, then $o_i(x) <_s d_i(x)$.\\
\texttt{MAL2}: Donated items are eventually unlocked; that is, if $d_i(x)$ occurs in a schedule $s$ following an operation $o_i(x)$, then $ou_i(x)$ is also in $s$ and $d_i(x) <_s ou_i(x)$.\\
\texttt{MAL3}: Transactions cannot hold conflicting locks simultaneously, unless one has donated the data item in question; that is, if $o_i(x)$ and $p_j(x), i \neq j$, are conflicting operations in a schedule $s$ and $o_i(x) <_s p_j(x)$, then either $ou_i(x) <_s pl_j(x)$, or $d_i(x)$ is also in $s$ and $d_i(x) <_s pl_j(x)$.\\

The terminology of wake, completely in wake, indebted also is on similar lines. Intuitively, if transaction $t_j$ locks a data item that has been donated and not yet unlocked by transaction $t_i$ , $i \neq j$, we say that $t_j$ is in the wake of $t_i$ . More formally, we have the following:
\begin{enumerate}
\item An operation $p_j(x)$ from transaction $t_j$ is in the wake of transaction $t_i$ , $i = j$, in the context of a schedule $s$ if $d_i(x) \in op(s)$ and $d_i(x) <_s p_j(x) <_s ou_i(x)$ for some operation $o_i(x)$ from $t_i$.
\item A transaction $t_j$ is in the wake of transaction $t_i$ if some operation from $t_j$ is in the wake of $t_i$ . Transaction $t_j$ is completely in the wake of $t_i$ if all of its operations are in the wake of $t_i$.
\item A transaction $t_j$ is indebted to transaction $t_i$ in a schedule $s$ if $o_i(x)$,$d_i(x)$,$p_j(x) \in op(s)$ such that $p_j(x)$ is in the wake of $t_i$ and either $o_i(x)$ and $p_j(x)$ are in conflict or some intervening operation $q_k(x)$ such that $d_i(x) <_s q_k(x) <_s p_j(x)$ is in conflict with both $o_i(x)$ and $p_j(x)$.

\end{enumerate}

\subsection{Shortcoming in AL}
\begin{equation*}
s_1=wl_1(a)w_1(a)d_1(a)rl_2(a)r_2(a)rl_2(b)r_2(b)ru_2(a)ru_2(b)c_2rl_1(b)r_1(b)wu_1(a)ru_1(b)c_1
\end{equation*}

$s_1$ is conflict serializable. But if $r_1(b)$ would be replaced by $w_1(b)$, $s_1$ would not be in CSR but still would be allowed by AL. So we had introduced AL4.\\ 
\texttt{AL4}: When a transaction $t_j$ is indebted to another transaction $t_i$ , $t_j$ must remain completely in the wake of $t_i$ until $t_i$ begins to unlock items. That is, for every operation $p_j(x)$ occurring in a schedule $s$, either $p_j(x)$ is in the wake of $t_i$ or there exists an unlock operation $ou_i(y)$ in $s$ such that $ou_i(y) <_s o_j(x)$.\\
So $s_1$ with either $r_1(b)$ or $w_1(b)$ is not passed by AL. $r_1(b)$ schedule is in CSR though. Thus a valid schedule is not passed through AL and hence poses an eminent shortcoming.

\subsection{Conclusion : AL $\subset$ MAL} 
In MAL, the conflicts are only $rw$ since only multiversion conflicts are considered. Thus consider two cases in the above $s_1$:
\begin{enumerate}
\item When $r_1(b)$, no problem is faced anyways.
\item When $w_1(b)$, a new version of $b$ is created and no new $rw$ conflict is created. Hence the schedule is still in MVCSR and hence also passed by MAL.
\end{enumerate} 
Hence MAL is more flexible and allows more concurrency than AL. 
Thus MAL4 is a more flexible version of AL4 in which the conflicts are of the form $rw$ instead of all $rw$, $wr$ and $ww$. 
Therefore it can be concluded that AL $\subset$ MAL.

\subsection{Need for MAL4}
\begin{equation*}
s=r_1(x)r_2(y)w_1(y)w_2(x)c_1c_2
\end{equation*}

In schedule $s$, $rw$ conflicts exist from $t_1$ to $t_2$ and $t_2$ to $t_1$. Hence the schedule is not in MVCSR. However it will get passed using MAL1-3 rules which should be prohibited. Therefore it is required to define another rule MAL4 to handle the problem.\\
\texttt{MAL4}: When a transaction $t_j$ is indebted ($rw$ conflicts only) to another transaction $t_i$ , $t_j$ must remain completely in the wake of $t_i$ until $t_i$ begins to unlock items. That is, for every operation $p_j(x)$ occurring in a schedule $s$, either $p_j(x)$ is in the wake of $t_i$ or there exists an unlock operation $ou_i(y)$ in $s$ such that $ou_i(y) <_s o_j(x)$.\\
We have now completely described the rules of MAL.

\section{Correctness}
\textbf{\emph{Gen(MAL) $\subset$ MVCSR}}\\ 
It essentially follows a standard argument, namely, that any MAL-generated history $s$ has an acyclic conflict graph. It can be shown that each edge of the form $t_i \rightarrow t_j$ in such a graph $G(s)$ is either a “wake edge,” indicating that $t_j$ is completely in the wake of $t_i$ , or a “crest edge,” indicating that $t_i$ unlocks some item before $t_j$ locks some item. In addition, for every path $t_1 \rightarrow \dots \rightarrow t_n$ in $G(s)$, there is either a wake edge from $t_1$ or $t_n$, or there exists some $t_k$ on the path such that there is a crest edge from $t_1$ to $t_k$ . These properties suffice to prove the claim.\\ Strict inclusion of MAL $\subset$ MVCSR has been shown later with an example.

\section{Extension of MV2PL}
We know that AL is an extension of 2PL where donation of locks is permitted. Long transactions hold onto locks until they commit and do not allow other transactions to execute. Similar problem can be observed in case of MV2PL as well. If a secondary small transaction needs to access a subset of data items which are currently locked by the primary transaction, read and write operation will get executed however commit will get delayed due to unavailability of the certify lock (certify lock is a type of lock that a transaction needs to acquire on all data items it has written to at the time of commit). Hence the secondary transaction will have to delay itself until the primary transaction releases all its locks.\\
If donation of locks is allowed in MV2PL then lock on certain data item can be donated to the secondary transaction which can commit without delaying itself by acquiring the certify lock. Handling of individual steps remains same as followed by MV2PL. Inclusion of donation of locks into MV2PL inspires the MAL scheduling protocol. In the next section we will infact see that \\MV2PL $\subset$ MAL.

\section{Comparison}
\subsection{AL $\subset$ MAL}
\begin{equation*}
s = r_1(x)r_2(z)r_3(z)w_2(x)c_2w_3(y)c_3r_1(y)c_1
\end{equation*}

Either $x$ or $y$ (or both) must be locked by $t_1$ between operations $r_1(x)$ and $r_1(y)$. By rule AL1, either $x$ or $y$ (or both) must be donated by $t_1$ for $w_2(x)$ and $w_3(y)$ to occur, so either $t_2$ or $t_3$ (or both) must be indebted to $t_1$. However, neither $r_2(z)$ nor $r_3(z)$ are allowed to be in the wake of $t_1$ if the latter is well formed, since $t_1$ later reads $z$. Hence either $t_2$ or $t_3$ violate rule AL4.\\
However as MAL allows donation of locks $t_1$ can donate lock to $t_2$ for certification and can commit. Hence $t_1$ need not acquire lock read lock on $y$ along with lock on $x$. Lock on $y$ can be obtained at read time.

\subsection{2PL $\subset$ MAL}
We know that 2PL$\subset$ AL as AL is a relaxed version of 2PL. Following the previous comparison 2PL$\subset$ MAL. Hence we can also conclude that 2PL$\subset$MAL.

\subsection{MV2PL $\subset$ MAL}
\begin{equation*}
s = r_1(x)w_2(x)w_2(y)c_2w_3(z)w_3(y)w_1(z)c_3c_1
\end{equation*}
Generating the output as per MV2PL rules, $r_1(x)w_2(x)w_2(y)$ will get executed by acquiring locks on respective data items. However $t_2$ cannot acquire certify lock on $x$ due to conflict with $rl_1(x)$ and will have to wait. $t_3$ will acquire $wl_3(z)$ and execute $w_3(z)$. Following this no transaction would proceed due to deadlock. $t_1$ can't acquire lock on $z$ due to conflict with $t_3$, $t_2$ cannot acquire certify lock on $x$ due to conflict with $t_1$ and $t_3$ cannot acquire write lock on $y$ due to conflict with $t_2$. Hence the schedule won't get accepted under MV2PL protocol.\\
In case of MAL $t_1$ can donate lock on $x$ to $t_2$ so that $t_2$ can commit using certify lock on $x$ and $y$. Following which $t_3$ can acquire write lock on $y$ and commit as well. At the end $t_1$ will commit by obtaining certify lock on $z$.

\subsection{2V2PL $\subset$ MAL}
2V2PL is just a special case of MV2PL where only two versions of a particular data item are allowed. Hence we conclude that 2V2PL $\subset$ MAL.

\subsection{Gen(MAL) $\subset$ MVCSR}
\begin{equation*}
s= r_1(x)r_1(y)w_2(x)w_2(y)w_1(y)c_1c_2
\end{equation*}
The $rw$ conflicts in schedule $s$ are from $t_1$ to $t_2$. The conflict is acyclic and the schedule is in MVCSR. But the MAL runs into a deadlock while scheduling $s$. $r_1(x)r_1(y)w_2(x)w_2(y)$ get executed by acquiring locks on respective data items. As $t_1$ cannot acquire write lock on $y$ due to conflict with $t_2$ the operation will get delayed. $t_1$ would have to donate its lock to $t_2$ for it certify write on $x$ and $y$. As per rule 1 of MAL, once a lock on a data item has been donated by a transaction, then that transaction cannot carry out any operation on that data item. Hence $w_1(y)$ will not get executed. Therefore the schedule cannot be generated by MAL.

\subsection{Gen(MAL) $\subset$ MVSR}
As MVSR $\subset$ MVCSR, using transitivity we can conclude that MAL $\subset$ MVSR.
\begin{figure}
\includegraphics[height=9cm,width=15cm]{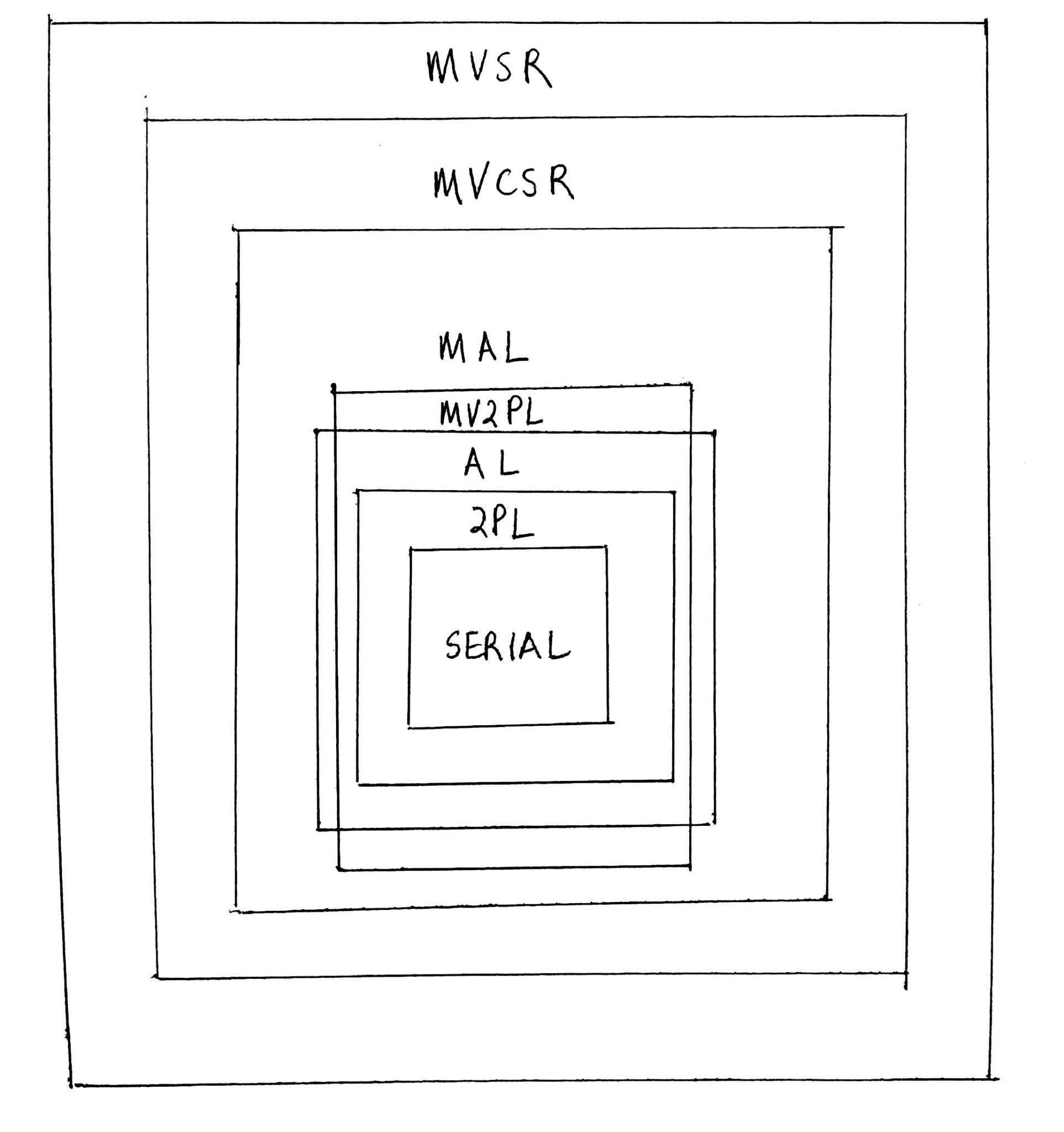}
\caption{Relationship diagram}
\end{figure}
\newpage
\section{Inclusion in Hybrid Protocols}
\textbf{MAL + MVTO}\\

\includegraphics[width=15cm,height=7cm]{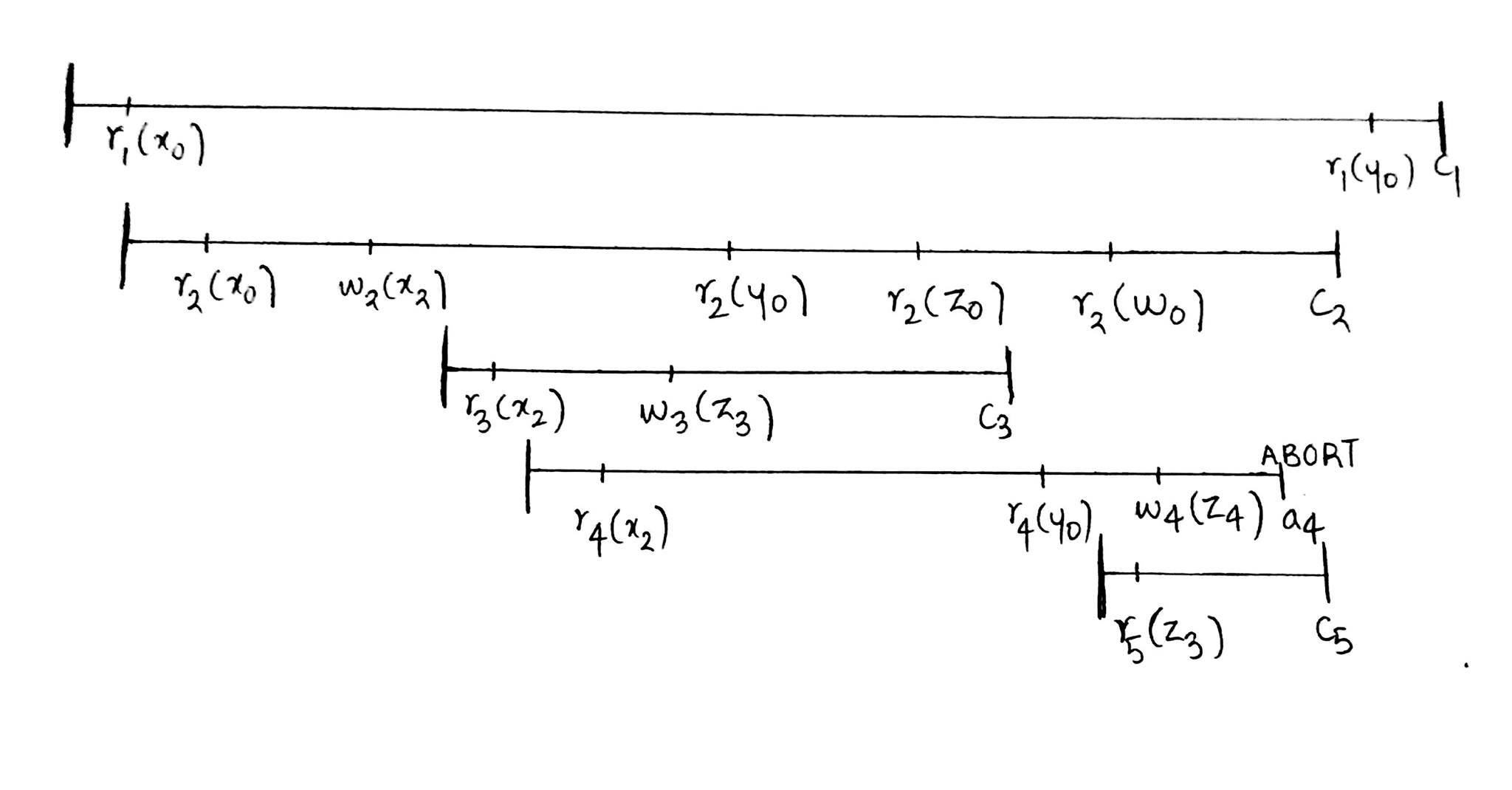}

Due to donations of locks, detection of aborted transactions of late writers can be done quickly saving both storage space and time.\\
If we know that a long transaction has only reads after a short span of the transaction time, it won't abort in MVTO (since aborts happen only due to write operations). In this case, $t_2$ is one such transaction. $t_3$ has a donated lock on $x$ from $t_2$. The altruism is predominant in the fact that a transaction can't commit until all transactions it has read from have committed. We change this. If we know $t_2$ has only reads after writing $x$, we know it won't abort. If $t_3$ reading from $t_2$ commits, $t_4$ is aborted since it has a late writer on $z$ ($t_5$ reads $z$ from $t_3$). $t_5$ is able to read $z$ from $t_3$ since it is committed; otherwise it would have to read from $z_0$ and hence $z_3$ and $z_4$ would have gone to waste due to $t_3$ waiting for $t_2$ to complete which would be a waste of space.\\ 
Thus MAL + MVTO is more successful than MVTO in this scenario. 

\section{Caveats of MAL}
\begin{enumerate}
\item Storage space would be required to store all versions of all variables.
\item This could be expensive if there are more \texttt{RW} transactions than \texttt{RO} transactions.
\item To avoid rollback, which would be very expensive considering the versions assigned, we should be pretty sure that there would not be any or very less number of aborts.
\end{enumerate}

\end{document}